\documentclass[aps,prd,onecolumn,12pt,preprintnumbers]{revtex4-1}

\usepackage{hyperref}

\usepackage{float}

\usepackage{tikz}

\usepackage{graphicx}
\usepackage{epstopdf}

\begin{document}

\preprint{HDP: 21 -- 04}

\title{Arithmetic of Spring Forces on the Banjo Bridge}

\author{David Politzer}

\email[]{politzer@theory.caltech.edu}

\homepage[]{http://www.its.caltech.edu/~politzer}

\altaffiliation{\footnotesize 452-48 Caltech, Pasadena CA 91125}
\altaffiliation{\newline \em \em \em 452-48 Caltech, Pasadena CA 91125}
\affiliation{California Institute of Technology}

\date{April 9, 2021}

\begin{figure}[h!]
\includegraphics[width=4.4in]{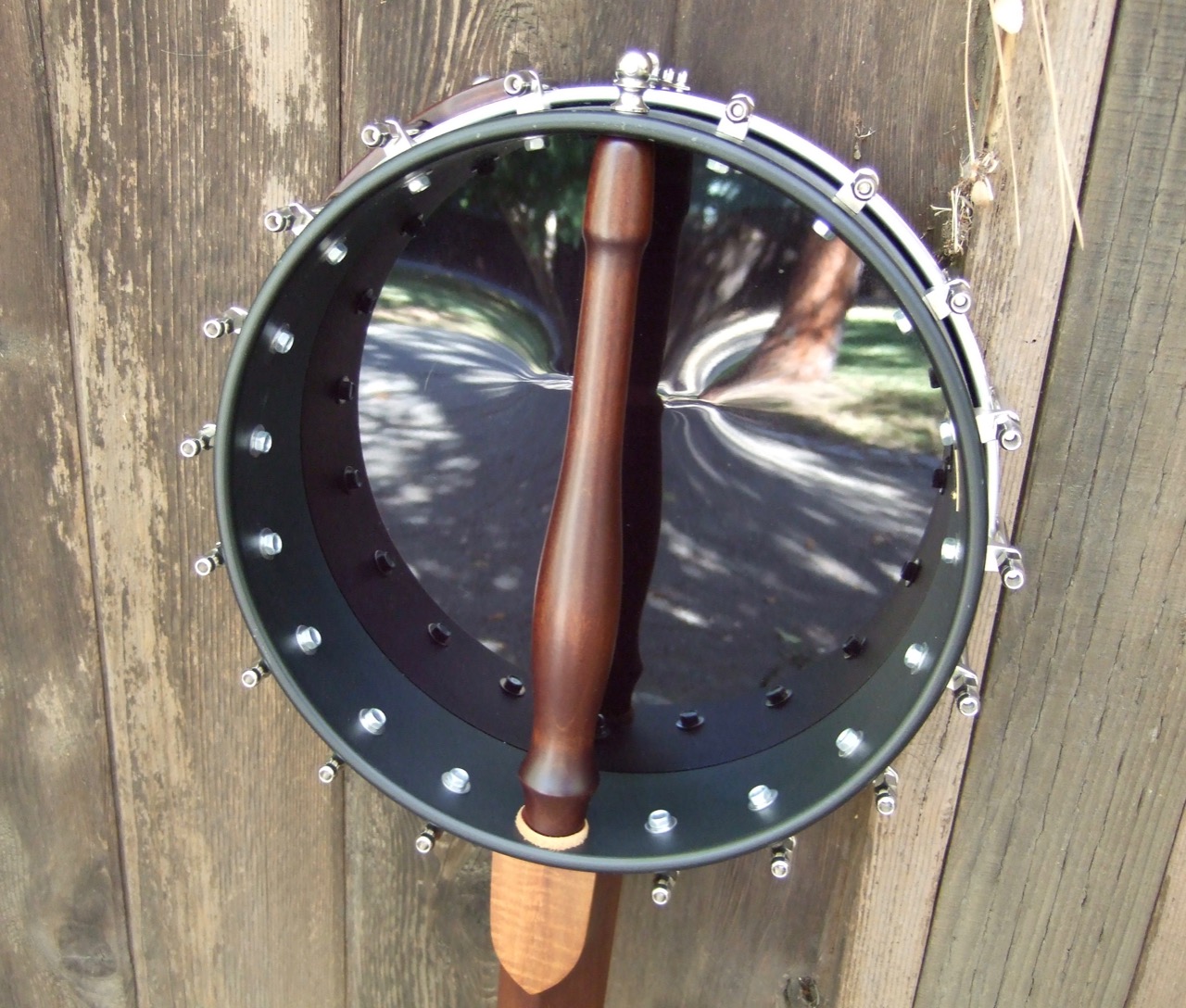}
\end{figure}

\begin{abstract}
Spring-like forces on the bridge are key to a banjo's characteristic voice.\cite{acoustics}  These are due to the tension in strings and head.  Conceptually distinct from the forces of waves in the strings and head that encode the underlying music, the spring-like forces impact the timbre of how those waves are converted to sound.  This note presents a simplified model that allows the head contribution to be calculated (mostly) with paper and pencil.  The key simplification is placing a circular bridge at the center of the head.  The resulting formulae show how design elements and player's adjustments can effect the sound.  The results also provide estimates of the magnitudes of the effects when evaluated with numerical values for Young's moduli and typical banjo set-up specs.   For steel strings and tight mylar head, the head contribution is about three times as large as that of the strings.
\end{abstract}

\maketitle{ {\centerline{\large \bf  Arithmetic of Spring Forces on the Banjo Bridge}}}

\section{Introduction}

In acoustic string instruments, strings transmit vibrations to a soundboard, which, in turn, produces sound in the air.  The strings and head have their own inherent resonances.  On banjos, the bridge is an important, {\it active} element in this process.  It has its own, qualitatively distinct resonances.  It functions as a resonant intermediary between strings and sound-producing head.  For the range of design parameters common to instruments identifiable as banjos, there is at least one main resonance and potentially a few more.  The main one involves the bridge moving rigidly as a whole and is the the focus of the present discussion.  The higher frequency ones involve bridge flexing.  Because these bridge resonances easily lose energy to the head, they all are very broad in frequency compared to the individual resonances of the strings themselves.  As a result, they produce regions in frequency of strong response to the initial string vibrations.\cite{acoustics}  In speech science and phonetics, such things are known as ``formants."  In the case at hand, they similarly are an important element in the characteristic voice of the instrument.  The main one is crucial to any version of banjo sound, while the others, all at higher frequencies, depend on details of bridge design and its interaction with its environs.

\begin{figure}[h!]
\includegraphics[width=4.0in]{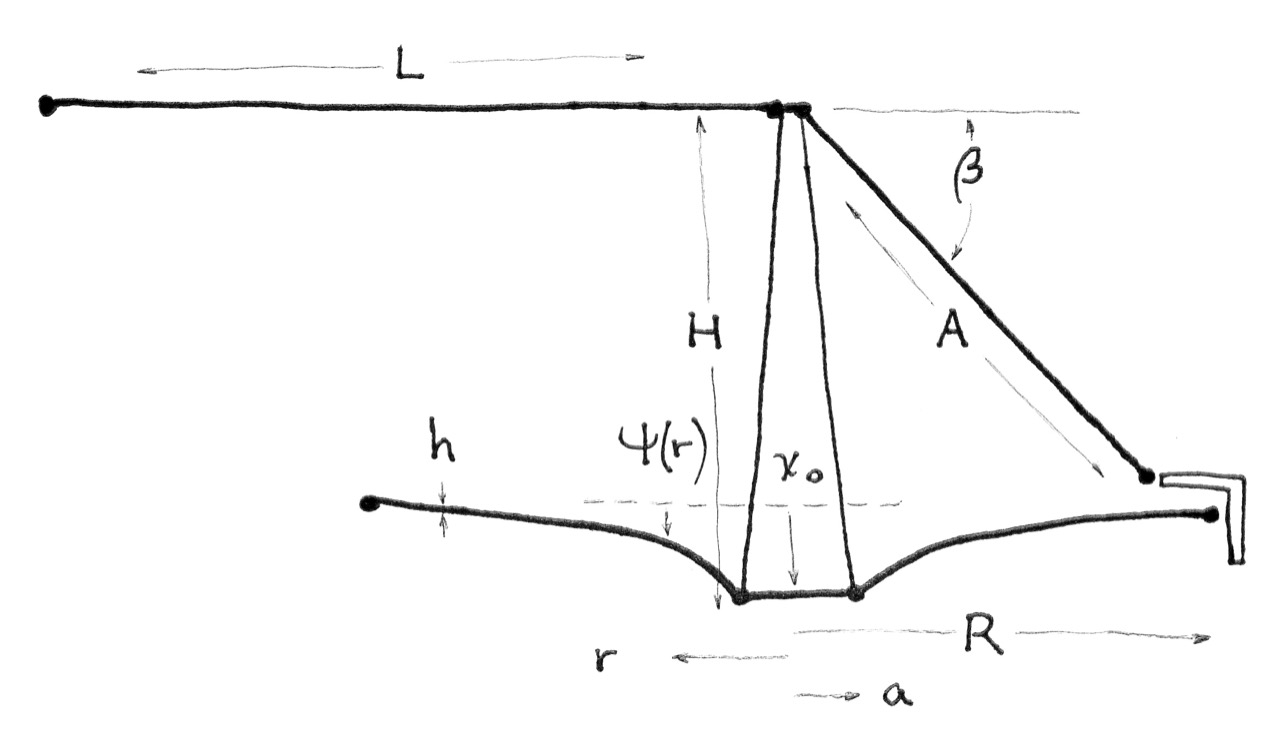}
\caption{The static tension of strings and head produces a linear return force on the bridge for small displacement from equilibrium --- as is clear from this idealized geometry (not to scale).}
\end{figure}


The amplitudes are all small enough that linearized physics gives a very satisfactory first approximation to the system.  For the main banjo formant, the bridge is a single mass moving under the influence of the Hooke's Law force provided by the pretensioned strings and head.  These forces are {\it in addition} to the forces due to waves on the strings and head acting on the bridge.  To isolate the main formant force, one need only consider a static displacement of the bridge with no waves present.  Banjo geometry produces a significant return force.  (See Fig.~1.)  The force is characterized by the Hooke spring constant in question.

This note offers a simplified model to illustrate how that works.  The final formulae show how the formant central frequency depends on variables at the builder's and player's disposal.  Realistic tensions and Young's modulus values can be plugged in to get rough estimates of the magnitudes that would arise in a more realistic geometry.  The string contributions are quoted in ref.~\cite{acoustics}.  The details are repeated here.  This note adds an analogous calculation for the head.

To reduce the head calculation to a paper-and-pencil exercise, the bridge foot is taken to be a circle at the center of the head.  Such an instrument could be built and would certainly sound like a banjo --- albeit a lousy one.  A more realistic model of the bridge, still within the framework of linear physics, would be a rigid bar located off-center.  But that is no longer soluble by hand.

According to ref.s~\cite{acoustics,bridge-hills}, there can be other formants that depend on the interplay of bridge flexing with the head underfoot.  But paper and pencil are insufficient to predict their frequencies and strengths.  You'd need some serious finite-element computer calculation of the type presented in ref.~\cite{acoustics} for the Deering 3-foot bridge.

\section{The Plan}

This is a linear vibration analysis.  Each piece of the physics involves an equilibrium configuration and small deviations.  In the equations presented here, the identity of the small items should be clear.  And the equations are approximate in that they represent the lowest order in the relevant parameters.  So all conclusions are approximations for small amplitude deviations from equilibrium.  As in most of musical acoustics, this is either good enough or a good place to start.

For simplicity, this discussion considers a single string.  The results for a 5-string banjo are not simply given by inserting a factor of 5 in appropriate places.  The bridge has strings placed along its length.  If all strings are not played simultaneously and identically, the net effect on a particular string's timbre due to the others will have a factor between 1 and 5 in some of those appropriate places.  This level of realism is left for the indeterminate future.

As sketched in Fig.~1, string and head tensions balance to establish an equilibrium position of the bridge.  Section \S III presents the equilibrium condition on the parameters.

The linear equation for the head itself is known as the ``ideal membrane equation."  Circularly symmetric boundary conditions (the rim) and a circular, central bridge allow a simple solution.   In that case, the equilibrium head displacement $\psi(\vec r)$ is a logarithm, i.e., $\psi(\vec r) \propto$ log $r/R$, where $\vec r$ is the position on the head, $r = |\vec r|$, and $R$ is the head radius, e.g., $11''$.  People familiar with logarithms usually remember three outstanding properties (i.e., the behaviors $r \to \infty$, $r \to 0$, and $r \approx 1$).  However, the behavior relevant to head shape is that from $a/R$ to 1 (where $a$ is the circular bridge radius), scaled to a small value at $r=a$.  That is shown in Fig.~2.  The photo on page 1 reflects the equilibrium shape of a standard banjo head.  Although the circular symmetry is absent, the curvature evidently increases dramatically in the vicinity of the bridge.

\begin{figure}[h!]
\includegraphics[width=5.5in]{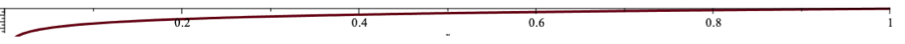}
\caption{A sketch of log {\small \it r/R} from $r/R = a/R=0.00001$ to $r/R=R/R=1$, scaled roughly to the actual head shape relevant to tensions in a real banjo.}
\end{figure}

When the bridge moves a small amount in the vertical direction (i.e., perpendicular to the head), the strings stretch and break angle increases, both by amounts proportional to the bridge displacement.  See Fig.~3.   Both of these give a linear increase in the string downward force on the bridge.

\begin{figure}[h!]
\includegraphics[width=2.83in]{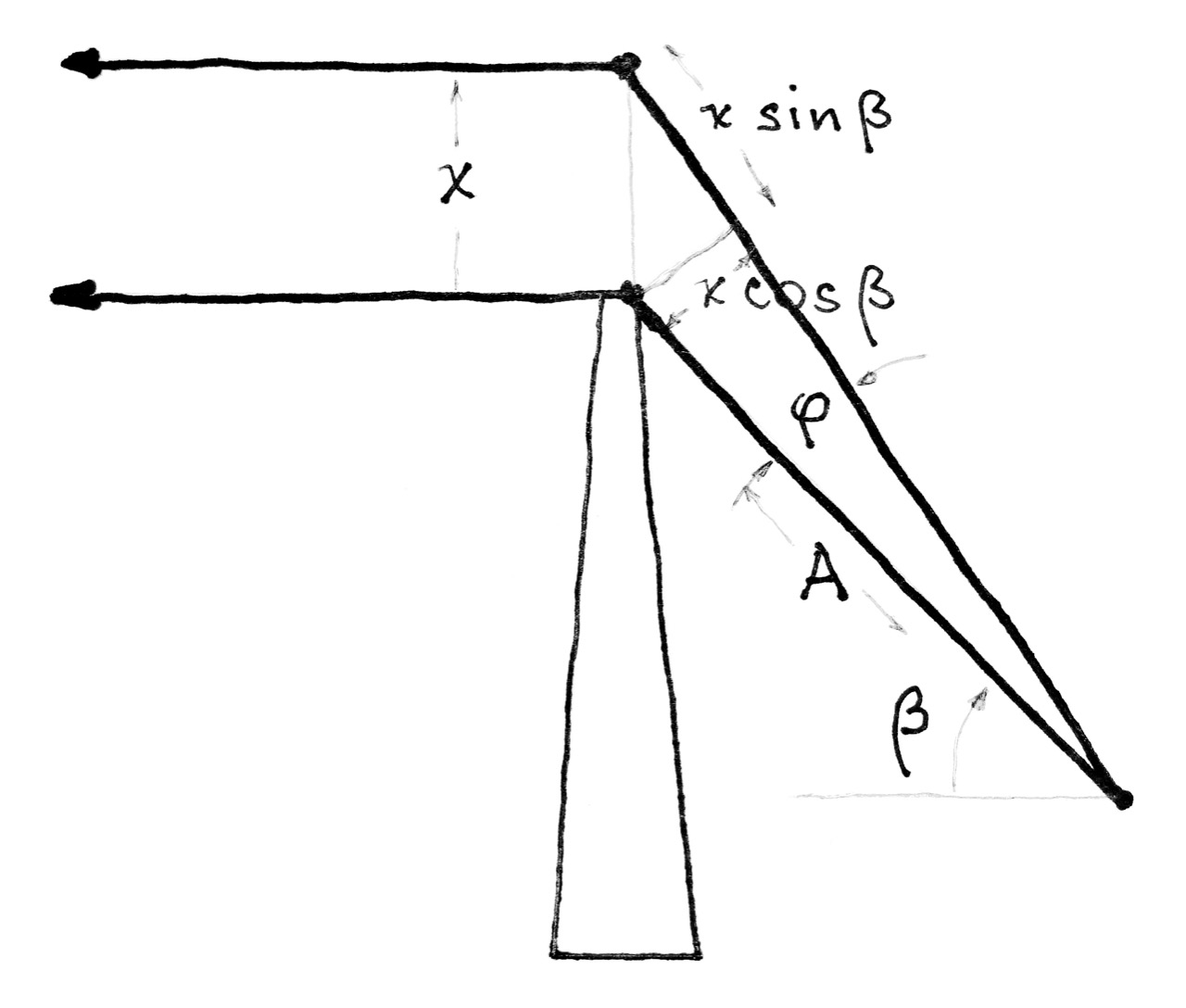}
\caption{Bridge displacement $x$ produces a stretch = $x$ sin$\beta$ and break angle increment $\phi = x$ cos$\beta$.}
\end{figure}

Vertical motion of the bridge also produces a first order stretch of the head in the radial direction and a change in its angle at the foot of the bridge.  That geometry is sketched in Fig.~4.  For a small vertical motion, these two head effects produce a force that is linear in the displacement that {\it adds} to the string force return force, i.e., they are of the {\it same} sign.

\begin{figure}[h!]
\includegraphics[width=4.7in]{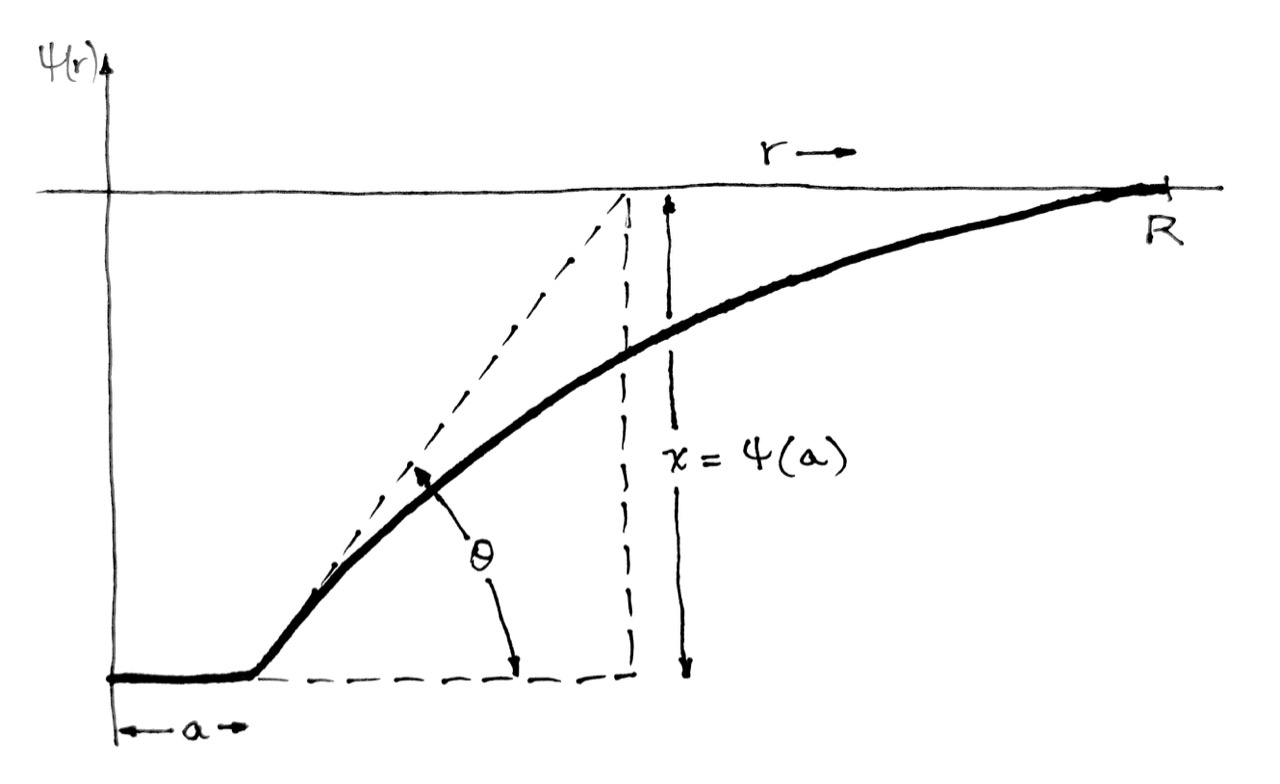}
\caption{Changing the bridge position, $x$, by $\Delta x$ produces  ${\cal O}(\Delta x)$ changes in both the the radial length of the head  and the break angle $\theta$. }
\end{figure}

\section{equilibrium}

At equilibrium, as sketched in Fig.~1, the downward force of the strings on the bridge is balanced by the upward force from the head.  The builder and player make a variety of design choices and adjustments that determine many of the parameters.  Then all other parameters relax to their equilibrium values.

\subsection{specified parameters}

String tension $T_s$ is the force along the string at any point.  The player sets it by choosing scale length $L$, string linear mass density $\rho$, and pitch frequency $f$.  String manufacturer D'Addario has a Web-based tool that produces tension values for choices of the other parameters\cite{D'Addario}.  Actually, they simply plug into the following formula:

\medskip
\centerline{$T_s = 4L^2f^2\rho$}

\noindent which follows from the standard equation for transverse waves on an ideal string.

The string break angle $\beta$ is determined principally by the bridge height and tailpiece design and possible adjustment.  $\beta$ and actually most of the parameters discussed here settle into their equilibrium values, responding to greater and lesser extents to the values of the others.  But $\beta$ is easy to measure once everything settles.  The symbols in the various equations presented here represent the equilibrium values.

Head tension $T_h$ is the force per unit length across any line on the head, assumed here to be a constant over the whole head.  Beginners often find the adjustment process mysterious but are delighted by the results when they learn to do it to their own taste.  Some banjo designs allow a considerable tension range provided by threaded hooks that pull the head material down over a rigid frame.  Even designs with the head fixed to the rim allow adjustment by choice of bridge height and tailpiece.  There are four methods used to ``get it right."  1) DrumDial is a gauge that measures deflection of the head near the rim.  The tension is a one-to-one, monotonic function of the dial reading.\cite{DrumDial}  (Few people care about the numerical force-per-unit-length values.)  2) You can measure the downward deflection of the head at the bridge (labeled $x_o$ in Fig.~1) by sighting across (if possible) or placing a rigid straight edge across the head.  3) You can tap and try to identify a pitch.\cite{tap}  And 4) you can listen for what sounds best.

\subsection{crude estimates}

The string-head equilibrium condition is simple in terms of the break angles $\beta$ and $\theta$ (Fig.s~1, 3, and 4): $T_s \times \sin \beta = T_h \times 2\pi a \sin \theta$.  There are two problems. $\theta$ is not addressed directly in banjo set-up.  And, in practice, it is hard to measure\cite{optical-lever} and somewhat ill-defined.  The theoretical sharp angle is rounded out by other forces beside the tension.  The same is true in principle for the strings, but the analogous region of rounding at the string break is tiny, and we appropriately quote the perceived angle just beyond that curvature.  Further away from the bridge, the string angle remains constant.  In contrast, even the theoretical linear modeling has the the head angle changing dramatically near the bridge --- as in Fig.~2.

The purpose of the central, circular bridge model is to offer an idea of rough magnitudes and how things work.  Contact with real banjo values is complicated by their having a distinctly non-circular bridge and its being far off center.  If one inspects closely in the manner suggested by the photo on page 1, it is clear that bridge feet (e.g., two or three of them) leave a complicated pattern of distortion in the head.  Nevertheless, here are some numbers to demonstrate that we're in the right ballpark.  The values for the Deering Eagle II studied in ref.~\cite{acoustics} are: $T_s \approx$ 27 kg or 270 N.  (That's for the sum of five strings.)  $T_h \approx 5.3$ kN/m.  The perimeter of the rectangle defined by the bridge feet is 0.17 m.  I'll take angles from a similarly set-up Deering Sierra: $\beta \approx 12^{\text{o}}$ and $\theta \approx 4.5^{\text{o}}$ .\cite{optical-lever}  Let $F_s$ and $F_h$ be the string and head equilibrium vertical forces on the bridge.  In the model and math below, I use $F_s = F_h$ to determine the equilibrium relations.  As a consistency and sanity check, here I can estimate them separately, albeit using a geometry and model that does not exactly match the banjo on which the input parameters were measured.  These numbers suggest $F_s \approx 55$ N, and $F_h \approx $ 71 N.  They're not exactly equal, but, given the huge range of input parameters as measured in SI units and substantial uncertainties, I regard this comparison as very encouraging.


\subsection{head deformation under central stress}

At equilibrium, a downward force --$F_h$ of the bridge on the head is matched by an equal and opposite force {\footnotesize+}$F_h$ of the head on the bridge.  Let $a$ be the radius of the bridge.  Let $r$ be the radial coordinate in the plane of the rim.  Let $\psi(r)$ be the vertical displacement of the head relative to flat.  Then

\centerline{$F_h = T_h$ $2\pi a \times$ \{slope of $\psi$ at $r$\}}

\centerline{$= T_h$ $2\pi a$ {\large${\partial \psi \over \partial r}$}{\Large $\vert$}{$_{r=a}$}}

\noindent
In fact, at equilibrium, this relation must hold, not just at $a$ but for all $r$, i.e.,

\centerline{$F_h = T_h$ $2\pi r$ {\large${\partial \psi \over \partial r}$}}

\noindent
This implies

\centerline{$\psi(r) =${\large${F_h \over 2\pi T_h}$}log{\large${r \over R}$}}

\noindent
where $R$ is the outer radius of the head.  (See note \cite{limits} for a subtlety here.)  This also tells us the equilibrium displacement of the bridge $x_o$:

\centerline{$x_o = \psi(a) =$ {\large${F_h \over 2\pi T_h}$}log{\large${a \over R}$}}

\subsection{string--head equilibrium condition}

The relation of string tension $T_s$ to its force $F_s$ on the bridge is clearly

\centerline{$F_s = - T_s \sin \beta$}

This can be combined with the result of $\S$IIIC to get an equation for the equilibrium position of the bridge, $x_o$:

\centerline{$x_o = ${\large ${1 \over 2 \pi}$}{\large ${T_s \over T_h}$} log{\large ${a \over R}$} $\sin \beta$}

\noindent
Again, this makes sense using the ref.~\cite{acoustics} values from the Eagle II for $T_s$, $T_h$, and $\beta$.  Taking $a$ to be the radius of a circle with the same circumference as the perimeter of the actual bridge footprint, gives $-\log{a \over R} \approx 1.65$.  Using the five string total for $T_s$ and this value of $a$ gives $x_o \approx 2.9$ mm or just under $1/8''$  --- an inordinately plausible number.

The logarithmic aspect of the head response deserves a qualitative comment.  The downward curvature of the log $r$ dependence means that, for a given $x
_o$, the slope at $x_o$ is steeper than were the radial dependence linear with $r$.  Hence, a smaller $x_o$ suffices to support the bridge with the log than would a linear shape.

\section{string stiffness}

Fig.~3 shows a bridge upward displacement $x$ produces a stretch $\Delta L = x \sin \beta$ and an increase in break angle $\phi = x \cos \beta / A$.

The strain $\epsilon = \Delta L / (L + A)$ produces a stress $\sigma = \Delta T = E_s \pi ({D \over 2})^2 \Delta L / (L + A)$, where $L+A$ is the total original length of string, $E_s$ is the string Young's modulus, and $D$ is the string diameter.  The change in force $\Delta F = \Delta T \sin \beta$.  Defining the stiffness constant $k_s$ by the relation $\Delta F = k_s x$, the string stretch contribution to the stiffness due to the string is

\medskip
\centerline{$k_{\text{string stretch}} =$ {\large ${E_s \pi (D/2)^2 \over L+A}$} $\sin^2 \beta$}
\medskip

The new total downward force at the original tension $T$ produced by the increase in break angle $\phi$ is $F + \Delta F = T \sin(\beta + \phi)$.  This implies $\Delta F = T \phi \cos \beta$, which gives the string angle contribution to the stiffness

\centerline{$k_{\text{string angle}} =$ {\large ${T \over A}$} $\cos^2 \beta$}

\section{head stiffness}

Fig.~4 and its caption suggest the analogous calculation that gives the head's contribution to the vertical stiffness at the bridge, $k_h$.  (Do not be confused by the cavalier treatment of signs.  The correct sign is always obvious.  Also $x$ in Fig.~4 is called $x_o$ here and elsewhere.)

Invert the solution for $\psi(a)$ in $\S$IIIC to find

\centerline{$F(x_o) = 2 \pi T_h x_o$ {\large ${1 \over \log{a \over R}}$}}

\medskip
Under $x_o \to x_o + \Delta x$, there are actually two ${\cal O}(\Delta x)$ contributions to $\Delta F$ (i.e., the partial derivatives): one from the explicit change in $x_o$ at fixed $T_h$ and the other from the stretch change in $T_h$ due to $\Delta x$, then evaluated with $x_o$. 
From the equation immediately above, the contribution to the stiffness from the explicit $\Delta x$ is

\centerline{$k_{\text{head - explicit}} = $ $2 \pi T_h$ {\large ${1 \over \log{a \over R}}$}}

The other contribution to the stiffness is due to the change in $T_h$.  That comes from the $\cal O$$(\Delta x)$ change in radial length along the head in going from $a$ to $R$ when $x_o$ is increased by $\Delta x$ at a fixed $T_h$.  For this, part it suffices to keep $x$ at $x_o$ and $T_h$ at its original value.  Then this new tension produces a new downward force to be evaluated with the original $x_o$.

The hard part is finding the radial stretch at fixed tension when $x_o$ is increased a bit. Mathematica produces an answer in fairly compact, closed form, even before evaluating the lowest relevant term in a power series.  That means there must be a way to do it by hand --- but I haven't found it.  So, I'll let Mathematica do it:

Let $L$ be the radial arc length along the head as $r$ goes from $a$ to $R$.  What we actually want is {\large ${\partial L \over \partial x}|_{x=x_o, \text{ fixed } T_h}$} .

\centerline{$L(a,R,T_o;x) = \int_a^R dr \sqrt{1 + ({\partial \psi \over \partial r})^2}$} 

\medskip
\noindent and, using the derived expressions for $\psi$ and $x$,

\centerline{\large ${\partial \psi \over \partial r} = {x \over \log{a \over R}} {1 \over r}$}
\smallskip
\noindent
Thus, the question of stretch, ${\partial L \over \partial x}$, is ``reduced to quadrature."  Mathematica offers the answer in closed form

\centerline{\large ${\partial L \over \partial x}|_{x=x_o, \text{ fixed } T_h} = {1 \over \log {a \over R}}\{\text{\small arcsinh} {z \over a} - \text{\small arcsinh} {z \over R}\}$}

\smallskip
\noindent where $z = x_o / \log {a \over R}$.  In practice, $x_o < a$ and $| \log {a \over R} | > 1$.  The first term in the expansion of arcsinh $y = y - y^3/3!$ + ... suffices to give the result:

\medskip
\centerline{{\Large ${\partial L \over \partial x}|$}$_{x=x_o, \text{fixed} T_h}$ {\Large $= {x_o \over (\log{a \over R})^2} [{1 \over a} - {1 \over R}] = {\Delta L \over \Delta x}$}}

\medskip
The strain $\epsilon = \Delta L / L$.  A sufficient first approximation to $L$ for this purpose is $L = R-a$.  That is the lowest order in the expansion of  the integral for $L$ in powers of $x$.  It's also obvious.

The increment in tension $\Delta T$ due to $\Delta L$ is

\centerline{$\Delta T =$ {\large ${1 \over 2 \pi a}$} $E_h {\cal A}$ {\large ${\Delta L  \over L}$}}

\noindent
Here ${\cal A} = 2 \pi a h$,  $h$ is the thickness of the head, and $E_h$ is the head Young's modulus.  So, the increment in the vertical head force on the bridge is

\centerline{$\Delta F =$ {\Large ${2 \pi x_o \over \log {a \over R}}$} $E_h h$ {\large ${\Delta L \over L}$}}

\medskip
\noindent $k_{\text{head stretch}}$ is identified from $\Delta F = k_{\text{head stretch}}\Delta x$.  The formulae above give

\medskip
\centerline{$k_{\text{head stretch}} =$ {\Large ${2 \pi E_h h \over (\log {a \over R})^3}{x_o^2 \over a R}$}}

\section{numerical estimates from Young's moduli and  tensions}

Ref.~\cite{acoustics} provides the following numbers for the Deering Eagle II (some already quoted in $\S$IIIB): for the sum of all five strings $T_s \approx 270$ N,  $L+A \approx 0.76$ m, $h \approx 1.8 \times 10^{-4}$ m, and $T_h = 5.3$ kN/m.  The perimeter of the rectangle defined by the bridge feet is 0.17 m, suggesting $a \simeq 2.7$ cm, i.e., for a circle with the same perimeter.  For steel, $E_s \approx 200$ GPa.  For a mylar head $0.007''$ thick, I found\cite{mylar} $E_h \approx 5.3$ GPa.


For the total string stiffness, I quote here the five-string approximate total.  For a normal shape bridge, the total would certainly be less but certainly more than 1/5, the value for a single string.

\centerline{$k_{\text{string stretch}} = 5.7$ kN/m} 
\centerline{    $k_{\text{string angle}} = 2.6$ kN/m}
\centerline{$k_{\text{head - explicit}} = 20$ kN/m}
\centerline{$k_{\text{head stretch}} = 2.9$ kN/m}

\medskip
\noindent That gives about $k_{\text{total}} \sim 31$ kN/m, with 73\% due to the head.

A rough estimate for $k_{\text{total}}$ for the Deering Eagle II from fitting the measured bridge admittance gives 20 kN/m.  (See Fig. 11b of ref.~\cite{acoustics}, paper \#2: {\it ...theoretical and numerical modeling}.)  Again, given the simplifications in doing the calculation and the enormous range of parameter numerical values that enter, this is an encouraging level of agreement.  In principle, $k_{\text{total}}$ can be measured directly.  (However, my home is not suitably instrumented in this safer-at-home era.)  In fact, the ``Compliance" discussed in ref.~\cite{acoustics} is $1/k_{\text{head explicit}}$.   That was done with no strings and $x_o = 0$, and it was measured as a function of $a$.  The numerical value for $a$ chosen for the estimates presented here is just at the onset of deviation from ideal membrane behavior.

\section{the resulting admittance}

Combining $k_{\text{total}}$ with the Eagle II bridge mass of 2.2 gm, we can imagine an oscillator frequency: $f_o = {1 \over 2 \pi} \sqrt{k_{\text{total}}/m_{\text{bridge}}} = 600$ Hz.  
Bridge motion is damped by outgoing waves on the head, whose calculation involves numerical evaluation including the Bessel (Hankel) functions that describe radial outgoing waves.\cite{acoustics}  If $\sigma_h$ is the head mass per unit area, i.e., $\sim 0.30$ kg/m$^2$, the head wave parameter combination {\large ${a \over m_h }$}$\sqrt{T_h \sigma_h} \approx 490$ Hz, sets the scale of the full-width-at-half-maximum.  The peak resonant frequency of a simple harmonic oscillator with linear, viscous damping (i.e., characterized by a single damping constant) is shifted down by the damping.  In the present situation, the bridge motion loses energy to the waves in the head, and the head admittance increases with frequency.  That admittance  dependence pushes the resonant peak higher in frequency.  Hence, the actual peak value is the resolution of the two opposing effects.

The point admittance at the bridge is a measure of the motion produced by an applied force.  It is typically presented as a function of frequency.  In normal playing, that force is provided by the vibrations of the strings.  And the admittance controls how much of those vibrations get onto the head to make sound.  It can be measured by tapping the bridge and measuring the motion.  Care is required to do a reasonable job.  And such measurements are reported in ref.~\cite{acoustics}.  Key ones are displayed in the top panel of Fig.~5.

Skudrzyk\cite{skud} pioneered a way to understand the trends and predict a smoothed version of the curves.  The idea goes roughly as follows.  The peaks occur at frequencies for which waves reflected off the edge return to the driving point in phase with the drive.  The dips are where the returning waves are $180^{\text{o}}$ out of phase.  The smooth curve is what happens if there are no return waves at all.  That occurs for a perfectly absorbing boundary or, more simply, for an infinite head.  It is simple to include a point bridge mass and stiffness, but the answer is in terms of the functions that describe outgoing membrane waves.  These are known as Hankel functions, and very little evaluation can be done by hand.  However, 
Mathematica knows Hankel functions.  The results for the parameters discussed in this note are shown in the lower panel of Fig.~5.  For both panels, the blue curves are the admittance of the head with no bridge or strings.  That serves as the primary damping of the mass/stiffness oscillation of the bridge and leads to the admittances in red.



\begin{figure}[h!]
\includegraphics[width=4.3in]{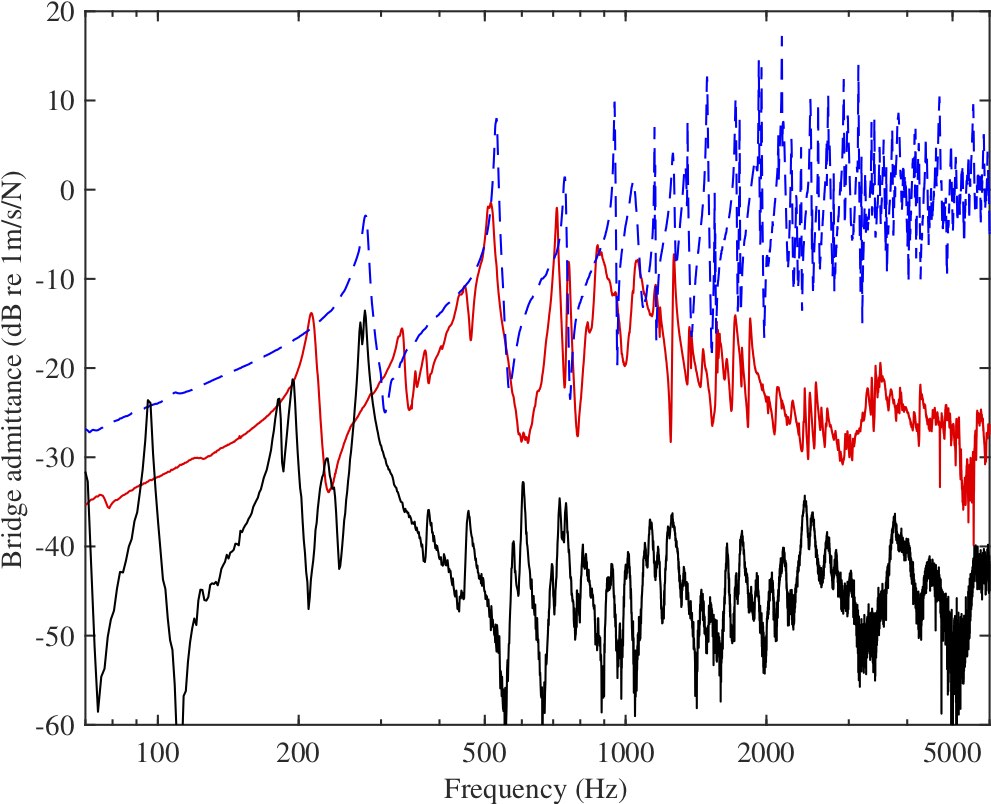}
\includegraphics[width=4.3in]{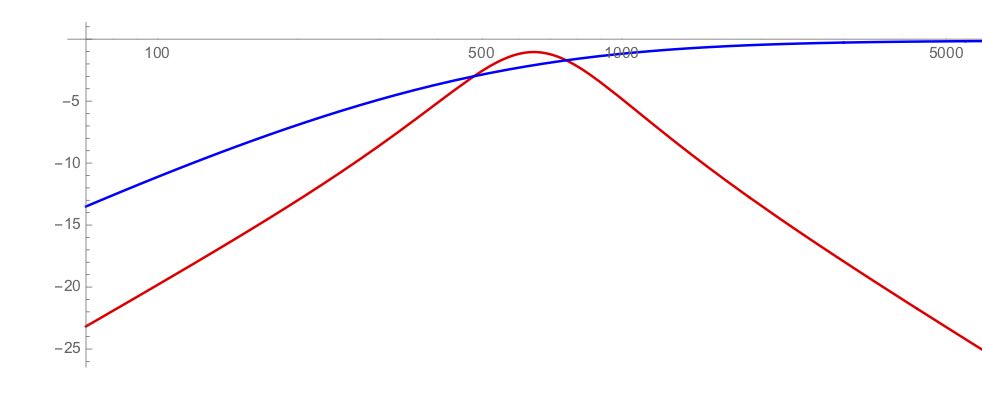}
\caption{upper figure from ref.~\cite{acoustics}:  {\footnotesize Measured bridge admittance at the location of the 1st string slot: blue is the head without bridge \& strings, red is with bridge and damped strings, and black is guitar with damped strings;}  lower figure: {\footnotesize calculated, averaged admittance (head alone in blue and including bridge and damped strings in red) with the model described above and parameters from the actual measured instrument.}}
\end{figure}


The upper graph includes an additional formant that begins around 3000 Hz.  In ref.~\cite{acoustics}, that is identified with flexing of the bridge and is not included in the model calculation of the lower graph.

The lower graph is computed using various parameters taken from the actual instrument measured for the upper graph.  However, the calculations use a bridge of the wrong shape and location, i.e., a cylinder at the center.  Also, the only energy ``loss" in the model is due to outgoing waves in the head.  There is no energy put into sound or heat.  Those are identified and measured in ref.~\cite{acoustics}.  The qualitative features of sound radiation from a finite-size membrane are essential to understanding the sustain of individual harmonics --- a feature that sets banjos apart from wood-topped instruments.  And in various places, at least some loss to heat is essential in getting good agreement of details of individual resonances between modeling and measuring.  But, overall, dissipation plays a minor role in understanding the physics.  The extent to which that's true is reflected again in the comparison presented in Fig.~5.


\section{summary} 

Inspired by the general success of applying the simplest models of strings and membrane to banjo acoustics, I extended that approach to estimating the head contribution to the stiffness.  A geometry with circular symmetry renders the model simple enough for evaluation by hand.  That is sufficient to identify what's big and what's little and how the main construction and adjustment choices impact the sound.  Of course, the impacts of these choices are no surprise.  They are  well-known elements of pickers' phenomenology.


\begin{thebibliography}{99}
\bibitem{DrumDial}
\noindent
\begin{figure}[h!]
\includegraphics[width=3.7in]{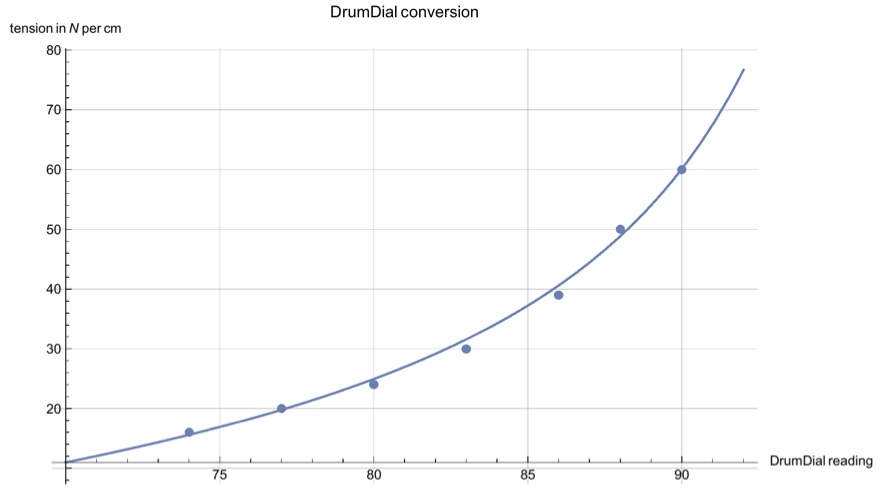}
\caption{\footnotesize The points are data provided by the manufacturer; the line is a plausible interpolation/extrapolation.}
\end{figure}
\bibitem{optical-lever}{\footnotesize My colleague Jim Woodhouse suggested using a laser pointer as an optical lever: a reflected distant spot jumps dramatically when the pointer aim is switched from under the bridge to the immediately adjoining head.  This works well on smooth mylar bottoms.  But it does not obviate the problem that the break pattern is actually very complicated around a three-foot bridge.  The photo on page 1 is actually an alternate implementation of the same idea.  A reflected image of graph paper, rather than tree and driveway, could be used to get a quantitative assessment of the surface orientation of the whole head simultaneously.}
\bibitem{limits}{\footnotesize In the formula presented, for any fixed $F_h$ and $T_h$, the magnitude of $\psi$ grows without bound as $r \to 0$.  As $|\psi|$ grows, the membrane has to stretch --- ultimately, a lot.  In any real, physical system, tension can no longer be the only relevant force.  Soap bubbles and balloons break when pricked.  With a banjo head, there will be significant forces from the stretching long before it breaks.  In practice, the bridge feet are likely large enough that tension dominates the behavior until right near the foot edge. One way of thinking about the abstract $r \to 0$ problem (for which an infinite amount of stretching is inevitable) is to consider $F_h \to 0$ as well.  Then, there is an obvious order of limits in which the membrane shape is, indeed, logarithmic down to $r=0$.  In the present application, it is assumed in all derivations and equations that the slope of the head is small enough that $\sin \theta \approx \tan \theta$.  That's a 6\% error at $20^{\text{o}}$, 1.5\% at $15^{\text{o}}$, and 0.3\% at $10^{\text{o}}$.}
\end{thebibliography}
\end{document}